# High precision binding energies from physics-informed machine learning

I. Bentley and J. Tedder
*Department of Physics, Florida Polytechnic University, Lakeland, Florida 33805, USA*

M. Gebran
*Department of Chemistry and Physics, Saint Mary's College, Notre Dame, Indiana 46556, USA*

A. Paul
*The Institute for Experiential AI, Northeastern University, Boston, Massachusetts 02115, USA*
*and CDNM, Brigham & Women's Hospital, Harvard Medical School, Boston, Massachusetts 02115, USA*

Twelve physics-informed machine learning models have been trained to model binding energy residuals. Our approach begins with determining the difference between measured experimental binding energies and three different mass models. Then four machine learning approaches are used to train on each energy difference. The most successful ML technique, both in interpolation and extrapolation, is the least squares boosted ensemble of trees. The best model resulting from that technique utilizes eight physical features to model the difference between experimental atomic binding energy values in AME 2012 and the Duflo Zuker mass model. This resulted in a model that fit the training data with a standard deviation of 17 keV and that has a standard deviation of 92 keV when compared all of the values in the AME 2020. The extrapolation capability of each model is discussed, and the accuracy of predicting new mass measurements has also been tested.

## I. INTRODUCTION

Theoretical binding energies can provide guidance for experimental measurements facilities working on the frontiers of science (e.g., at CARIBU [1], at JYFLTRAP [2], and at FRIB [3,4]). Further from stability, theoretical binding energies play a critical role in astrophysically relevant calculations, for example, regarding the r-process (e.g., Refs. [5] and [6]). The r-process occurs near the proton drip line, where experimental measurements have not been thoroughly explored. For these cases and others, high precision theoretical binding energies are sought out.

There has been a long history of binding energy models dating back to Refs. [7] and [[8]. The best contemporary atomic binding energies models typically reproduce atomic mass evaluation (AME) experimental values for $N$ □ 8 and $Z$ □ 8, with standard deviations, denoted as $\sigma$ , ranging from about 200 to 800 keV (for a summary discussion, see Ref. [9] and the references therein, or Ref. [10]).

Recently, the use of machine learning (ML) techniques informed by a dozen or so physical features have been able to achieve models of binding energy using neural networks [11,12], or support vector machines (SVMs) and Gaussian process regression (GPR), which also result in binding energy models in this range of accuracy [13]. A best fit value as low as $\sigma$ = 128 keV has been achieved using a kernel ridge regression that included odd-even effects [14]. The technique of training ML models informed by physical features has also been adapted to predict $\alpha$ decay [15], quadrupole deformations [16], and B(E2) values [17].

In this manuscript we present a methodology to predict the binding energy involving four different ML approaches: SVM, GPR, neural networks, and ensemble of trees. In general the SVM effectiveness is particularly evident when handling nonlinear regression challenges [18], often delivering superior outcomes to other techniques [19]. GPR is similarly kernel-based, but it results in probabilistic models that are nonparametric [20,21]. Artificial neural networks utilize hidden layers that mimic connections between neurons in the brain through activation functions and weights, allowing them the ability to determine nonlinear relations [22]. Ensemble of trees algorithms are widely utilized ML techniques for analyzing tabular data sets [23]. These ML approaches were used to train models on a subset of the AME 2012 data [24] based on groups of physical features.

Instead of modeling the binding energy directly, our approach consists of training ML models on the residual difference between the experimental values and one of three different mass models. The mass models used are two simple semiempirical models and the Duflo Zuker [25] mass model. Section II discusses the binding energy models and data sets used to train and test. This alleviates the burden on the ML models of learning the underlying distribution of the binding energies and allows for the accomplishment of a potentially easier task of predicting just the difference in the binding energies. In essence, once the background of "theoretical prediction" is removed, it becomes easier to model the "signal," which is the difference between the experimental measurement and the theoretical predictions. In other words, what the ML model is left to learn from binding energy differences are the terms dependent on the physical features that are not explicitly accounted for in the theoretical models, i.e., higher-order effects in terms of the physical features. Furthermore, we explain the predictions of the models using Shapley values, a measure of feature importance derived from cooperative game theory.

Section III discusses the four ML approaches used and the relevant hyperparameters that have been optimized in each. Section IV discusses the process used to determine which physical features are critical to the ML model training. Section V discusses the process used to determine the best models and corresponding model fit metric values. This section also contains a discussion comparing the predictive power and extrapolation limitations of the different ML approaches.

## II. DETERMINING THE TRAINING AND TESTING DATA

When training ML models the preprocessing of the data is essential, as it makes the learning task easier for the ML models. In this work the preprocessing of the data comes in the form of removing one of three mass models from the experimental values.

### A. Five-coefficient model of binding energy

The process of creating a macroscopic mass model usually begins by treating the nucleus as a liquid drop of charged nuclear matter comprised of protons and neutrons. Modern versions of this technique can achieve models on the order of 400 keV [26].

In that spirit, a simple five-coefficient binding energy model originally introduced as Eq. (3) in Ref. [27] has been chosen, which is of the following form:

$$\begin{aligned} B_{\mathrm{LD5}} = &(a_v A + a_s A^{2/3})[1 + \kappa T_Z(T_Z+1)A^{-2}] \\ &+ [a_c Z(Z-1) + \Delta]A^{-1/3}, \end{aligned} \tag{1}$$

where the neutron number ($N$), proton number ($Z$), the mass number ($A$), and the isospin projection [$T_Z = (N-Z)/2$] are used. The pairing contribution used is $\Delta = +a_p$ if the nucleus is even-even, $\Delta = -a_p$ if odd-odd, and is $\Delta = 0$ otherwise.

This expression contains the standard liquid droplet model surface and volume considerations. The $Z(Z-1)$ expansion results from the semiclassical treatment of the protons accounting for the interaction of each proton with all other protons. The expansion of the form $T_Z(T_Z+1)$ is consistent with experimental observations [28] and results from isovector pairing calculations in the strong pairing limit [29].

The coefficients $a_v = 15.79$ MeV, $a_s = -18.12$ MeV, $\kappa = -7.18$, $a_c = -0.7147$ MeV, and $a_p = 5.49$ MeV were chosen from the fit in Ref. [27], and when compared to the AME 2012, this results in a standard deviation of 2.662 MeV.

The difference between the experimental binding energy and the theoretical liquid drop binding energy,

$$\Delta B_{\mathrm{LD5}} = B_{\mathrm{expt.}} - B_{\mathrm{LD5}}, \tag{2}$$

will be the first of three residuals that ML models will be trained to match.

### B. Six-coefficient model of binding energy

The difference $\Delta B_{\mathrm{LD5}}$ is demonstrated in Fig. 1(a). In an attempt to model the remaining features, which often peak at magic numbers of protons and neutrons, and peak even higher for doubly magic nuclei, the following variables have been used:

$$\nu = \frac{2N - N_{\mathrm{max}} - N_{\mathrm{min}}}{N_{\mathrm{max}} - N_{\mathrm{min}}} \tag{3}$$

and

$$\zeta = \frac{2Z - Z_{\mathrm{max}} - Z_{\mathrm{min}}}{Z_{\mathrm{max}} - Z_{\mathrm{min}}}, \tag{4}$$

where the minimum and maximum values are defined by the nearest magic numbers. This results in $\nu$ and $\zeta$, beginning with values that increase linearly from -1 when adding valence nucleons above a closed shell. It is 0 when midshell, and 1 when the next shell is closed as discussed in detail in Ref. [30]. In this analysis the magic numbers chosen are

$$N_{\mathrm{min/max}} = [2, 8, 20, 28, 50, 82, 126, 196] \tag{5}$$

and

$$Z_{\mathrm{min/max}} = [2, 8, 20, 28, 50, 82, 114, 124], \tag{6}$$

as motivated by gaps in Nilsson levels from Ref. [31].

To recreate the parabolas that peak for magic numbers of protons and neutrons demonstrated in Fig. 1(a), the $\nu$ and $\zeta$ are added to create a sixth semiempirical term that can be written as

$$B_{\mathrm{LD6}} = B_{\mathrm{LD5}} + a_{\mathrm{Shell}}(\nu^2 + \zeta^2)^2. \tag{7}$$

This function of $\nu$ and $\zeta$ peaks, with a value of 4 for doubly magic nuclei, is one when there is one magic number and the other nucleon is midshell, or is zero if both the protons and neutrons are midshell. The corresponding best fit has determined the new coefficient to be $a_{\mathrm{Shell}} = 2.77$ MeV.

The second residual $\Delta B_{\mathrm{LD6}}$ is the difference:

$$\Delta B_{\mathrm{LD6}} = B_{\mathrm{expt.}} - B_{\mathrm{LD6}}. \tag{8}$$

The inclusion of this $\nu$- and $\zeta$-based term has reduced the standard deviation between the model and the AME 2012 experimental values by approximately 1 MeV to 1.667 MeV. This has been included in Fig. 1(b).

### C. Duflo Zuker

For the final residual, we utilize a well-known and trusted formula as opposed to attempting to further refine our own semiempirical models.

The Duflo Zuker (DZ) micromacro mass formula has been chosen as the third option from which to calculate the binding energy residuals. The DZ model using 28 parameters was published in 1995, and it had an accuracy of 375 keV for the 1751 nuclides with measured masses known at the time [25]. In the nearly 30 years since this model was introduced, it has been proven to provide an excellent binding energy estimate.

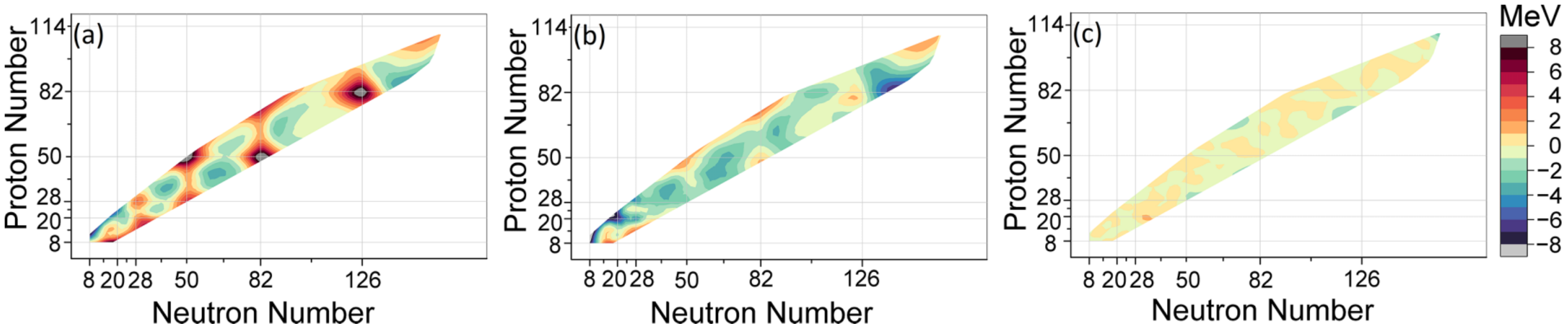

FIG. 1. Binding energy difference, (a) $\Delta B_{\mathrm{LD5}}$, (b) $\Delta B_{\mathrm{LD6}}$, and (c) $\Delta B_{\mathrm{DZ}}$, between the three theoretical models of interest, experimental measurements from AME 2012 [24].

Again, the residual used is the difference of the theoretical binding energy removed from the experimental binding energy:

$$\Delta B_{\mathrm{DZ}} = B_{\mathrm{expt.}} - B_{\mathrm{DZ}}. \tag{9}$$

The resulting values are displayed in Fig. 1(c).

### D. Nuclides used in the training and testing sets

These three binding energy differences have been chosen to represent (1) a crude model that does not consider shell structure, (2) another that does consider shell structure in a global manner, and (3) a more sophisticated and trusted model.

We find it critical to not train on the same data as is being tested. The training of all models discussed in this manuscript is based on the AME 2012 data [24]. The AME 2020 data [32] was used to create an independent testing set where none of the same nuclides were used.

The full AME 2012 contains experimental binding energy values for 2353 nuclides. In our procedure, some were removed and reserved for the training set. There are 17 nuclides that have experimentally measured values in AME 2012 but that are not found in the AME 2020. Those were removed from the training set. Additionally, 57 binding energy measurements changed by more than 100 keV when comparing AME 2020 with AME 2012. Those were also not included in the training set. Lastly, one out of every seven of the remaining nuclides (326 total) were also removed from the training set and were reserved for the test set. This last group was included, so the test set can be representative of the entire set. In total, 400 nuclei that are in the AME 2012 data set were not used to train.

In addition to the 57 nuclides with substantially changed binding energy values and the 326 corresponding to every seventh nuclide, there are 121 nuclides with new mass measurements in the AME 2020. All combined, an independent set of 504 nuclides from AME 2020 were used to test the accuracy values from the models. Figure 2 demonstrates the testing set nuclides and the locations and reasoning for the testing set nuclides. This represents approximately 20% of the total AME 2020 set.

Table I has been included to provide some perspective on how good each model is initially for the training, test, and full AME data sets. The number of nuclei and the mean Experimental Uncertainty ($\overline{EU}$) values have also been included for each data set. The value of $\overline{EU} = 20$ keV is important because all models will be trained on a data set that has average uncertainties at this level.

## III. MACHINE LEARNING APPROACHES

Four ML approaches have been used to train models for each of the differences: $\Delta B_{\mathrm{LD5}}$, $\Delta B_{\mathrm{LD6}}$, and $\Delta B_{\mathrm{DZ}}$. To protect against overfitting, the training involved a fivefold crossvalidation scheme, where the training data is partitioned and accuracy is estimated for each fold. The model hyperparameters were determined using a Bayesian optimization routine with the *expected improvement per second plus* acquisition function. The optimization process for each model was run for 250 iterations, although convergence often occurred within the first 50 iterations.

### A. Support vector machines

We have implemented a variant of the SVM [33], the support vector machine regression (SVMR) [34], as one of our machine learning approaches. The success of an SVM-based model in these scenarios hinges on the use of the kernel trick, a powerful technique for addressing data that is not linearly separable in its original feature space. By using the kernel trick, SVM implicitly transforms the data into a higher-dimensional space where linear separability can be achieved [35]. This transformation is made possible through a kernel (covariance) function, which computes the dot product between data points in this higher-dimensional space without explicitly calculating the transformed features. As a result, SVM maps input data into a higher-dimensional space using kernel functions, enabling it to capture nonlinear relationships.

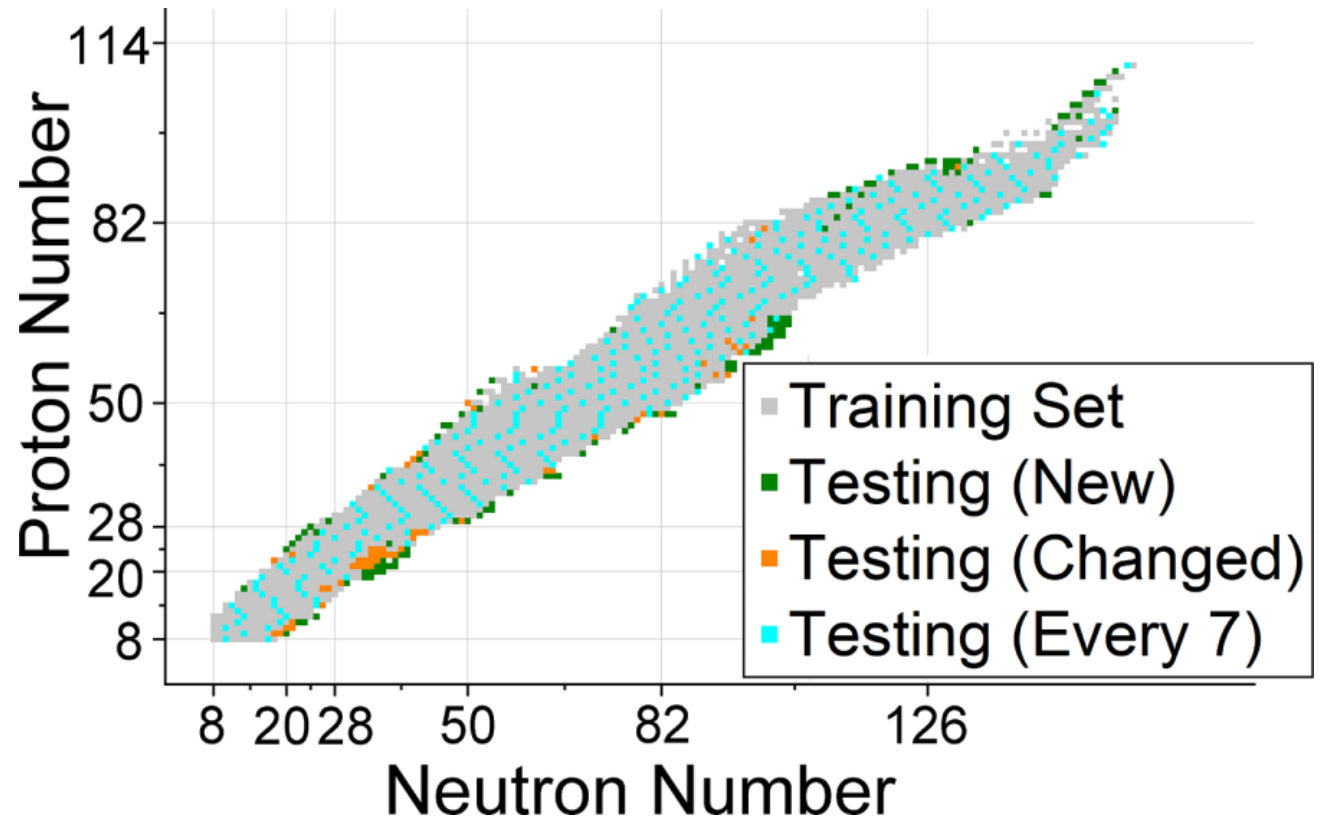

FIG. 2. The training set comprised of AME 2012 data along with the three constitutes of the testing set from AME 2020, specifically new values, changed values, and the selection for one out of every seven throughout.

TABLE I. Data set, size, mean experimental uncertainty, and model standard deviation comparison.

| Set<br>Number | Train 2012<br>1953 | All 2012<br>2353 | Test 2020<br>504 | All 2020<br>2457 |
|---|---|---|---|---|
| $\overline{EU}$ | 0.020 MeV | 0.026 MeV | 0.044 MeV | 0.023 MeV |
| $\sigma_{\Delta B_{\mathrm{LD5}}}$ | 2.638 MeV | 2.662 MeV | 2.773 MeV | 2.666 MeV |
| $\sigma_{\Delta B_{\mathrm{LD6}}}$ | 1.672 MeV | 1.677 MeV | 1.779 MeV | 1.696 MeV |
| $\sigma_{\Delta B_{\mathrm{DZ}}}$ | 0.384 MeV | 0.393 MeV | 0.561 MeV | 0.427 MeV |

The SVMR approach is specifically crafted for regression tasks, offering a distinctive method for predicting continuous values by applying the principles of SVMs. Unlike classification-based techniques, SVMR aims to identify a hyperplane that best models the training data while minimizing prediction errors. A key element of SVMR is its reliance on support vectors, which are the crucial data points located nearest to the hyperplane's boundaries. The model strives to position as many data points as possible within this optimal hyperplane, adhering to a defined tolerance margin while also managing instances where data points lie outside the boundary. This introduces a hyperparameter, which determines the hyperplane's width.

Similar to Ref. [13], we have determined that the Gaussian kernel of the form

$$K_G(x_i, x_j) = e^{-\gamma r^2} \tag{10}$$

generates the most reliable models. Here $x_i$ and $x_j$ represent two data points, and $r$ is the Euclidean distance between the two points:

$$r = ||x_i - x_j|| = \sqrt{(x_i - x_j)^T (x_i - x_j)}, \tag{11}$$

and $\gamma$ is the kernel coefficient.

The optimized hyperparameters are the kernel scale, the box constraint (labeled $C$) which controls the penalty imposed on observations with large residuals, and the $\epsilon$ values that govern the margin of tolerance.

### B. Gaussian process regression

GPR works by modeling a Gaussian distribution over possible functions. The algorithm begins by defining a prior distribution over these functions, where the function values at the input points follow a Gaussian distribution [21]. This prior is based on a mean function and a kernel function, which sets the initial assumptions.

As training data is introduced, the prior is updated to form a posterior distribution by applying Bayes's theorem. This update integrates the observed data, adjusting the model's understanding of the underlying function. The posterior distribution then allows for predictions at new data points, offering not just a predicted mean but also a measure of uncertainty, which is especially valuable for making decisions in uncertain environments.

The strength of GPR lies not only in its predictive accuracy but also in its capacity to provide detailed insights into the confidence of those predictions. This is accomplished through its uncertainty quantification, which offers a probabilistic assessment of how reliable each prediction is. By accurately measuring the uncertainty associated with each outcome, GPR significantly improves decision-making processes. This combination of delivering accurate predictions while simultaneously evaluating their reliability makes GPR a powerful tool.

We have trained our GPR models using the isotropic Matern 5/2 kernel function, which also involves $r$ from Eq. (11), of the form

$$K_{M5/2}(x_j, x_k) = \sigma_f^2 \big(1 + \sqrt{5} r/\sigma_l + 5r^2/3\sigma_l^2\big) e^{-\sqrt{5} r/\sigma_l}, \tag{12}$$

where $\sigma_l$ is the characteristic length scale and $\sigma_f$ is the signal noise standard deviation.

The optimized hyperparameters in this approach are the kernel scale, the type of the basis function used, either zero, constant, or linear, and $\sigma_f$.

### C. Neural networks

Neural networks are differentiable universal function approximators that have nonlinearity introduced through activation functions that act on an affine transformation of the input using learnable weights and biases [36,37]. Implemented as multilayer perceptrons, they are nested functions represented by hidden layers that are fully connected. The complexity (or expressivity) of a neural network relies on the number of hidden layers, representing the depth of nesting, and the width of the hidden layers, representing the number of activations summed over in each layer.

Our approach involves fully connected neural network (FCNN) models with one, two, or three hidden layers. The options tested for the number of nodes on each hidden layer were [300], [1000], [10 10], [100 100], [200 200], [300 300], [10 10 10], [100 100 100], [200 200 200], [300 300 300], and [400 400 400]. Additionally, three activation functions were tested: ReLU, sigmoid, and tanh. An $L2$ regularization was used, requiring the optimization of a hyperparameter corresponding to the regularization strength (denoted as $\lambda$).

### D. Ensemble of trees

Ensemble of trees–based algorithms utilize multiple base decision trees, each trained on a separate subset of the data. Notable examples of tree-ensemble algorithms include random forests [23], gradient boosted decision trees [38]), or the least squares boosting (LSBET) [39].

Among these ensemble algorithms, we have determined that LSBET is the preferred technique for modeling binding energy residuals. It effectively leverages the strengths of boosting and decision trees to improve predictive accuracy in regression scenarios. The process begins by initializing the first trees, then predicting the residuals, and then based on the prediction, a step is taken which is scaled by the learning rate (denoted as $\eta$). Finally, the squared error between the predicted and true values is minimized. Overfitting can be protected against by using a low learning rate, regulating the maximum depth of the trees, and limiting the number of boosting rounds.

In our tests, we varied the number of learners to be 10, 100, 300, 1000, 1500, 3000, 6000, 9000, and 12000 for these calculations, and the optimized hyperparameters are the minimum leaf size and the learning rate.

## IV. PHYSICAL FEATURE SELECTION

Our approach is to determine binding energies based on ML approaches that have been trained on energy residuals coming from each of the $\Delta B$ sets. This differs from the prior works, specifically Refs. [11] and [13], where the binding energy values are used directly. Training ML models on the residuals may lead to more accurate models, and it may reduce the number of physical features used.

The list of physical features used contains two pairs of quantities directly related to the number of particles in the nucleus, specifically, $N$, $Z$, $T_Z$, and $A$.

The two linear shell structure parameters from the added term in $B_{\mathrm{LD6}}$, namely, $\nu$ and $\zeta$, have also been added. This is similar to the inclusion of the valence neutron ($V_N = N - N_{\mathrm{min}}$) and valence proton ($V_Z = Z - Z_{\mathrm{min}}$) numbers using Eqs. (5) and (6), and Casten's $P$ factor from Ref. [40] [where $P = (V_N V_Z)/(V_N + V_Z)$], used in other works.

The spherical subshell number for the last occupied neutron and proton, denoted as $N_S$ and $Z_S$, respectively, has also been included. This is similar to the use of shell terms in Refs. [11] and [13], but in this case, the subshell count was used as opposed to the major shell.

If, for example, $N = 9-14$, then according to the spherical Nilsson model (from the Nuclear Structure section of Ref. [31]), the valence neutron will be on $1d_{5/2}$, by counting up from zero $N_S = 4$. If $N = 15-16$, then $2s_{1/2}$ will be occupied and $N_S = 5$. If $N = 17-20$, then $1d_{3/2}$ will be occupied and $N_S = 6$, and so on. Using the subshell as opposed to the major shell number provides the ML models more flexibility to train on a wider variety of features. If the major shell is an important feature, then, for example, $N_S = 4$, 5, and 6 can be treated identically and effectively grouped within the model, corresponding so that it behaves as one large oscillator shell.

The fifth group of parameters included to provide modifications resulting from even-odd mass staggering are Boolean values accounting for whether the number of nucleons is even or odd. These are written as $N_E$ and $Z_E$, and the value is 1 if the respective neutron or proton number is even, or 0 if it is odd.

The five pairs of features could be combined as $2^5 - 1 = 31$ possible groups. A preliminary set of calculations was made for each ML approach on each $\Delta B$ value where all ten physical features were included. A subsequent analysis of those preliminary calculations using Shapley values allowed for restrictions on the number of groups to be determined.

TABLE II. Physical features used with ML models.

| Feature group | Number of features | Physical features |
|---|---|---|
| 1 | 2 | $N, Z$ |
| 2 | 6 | $N, Z, T_Z, A, \nu, \zeta$ |
| 3 | 6 | $N, Z, T_Z, A, N_S, Z_S$ |
| 4 | 8 | $N, Z, T_Z, A, \nu, \zeta, N_S, Z_S$ |
| 5 | 8 | $N, Z, T_Z, A, \nu, \zeta, N_E, Z_E$ |
| 6 | 8 | $N, Z, T_Z, A, N_S, Z_S, N_E, Z_E$ |
| 7 | 10 | $N, Z, T_Z, A, \nu, \zeta, N_S, Z_S, N_E, Z_E$ |

Shapley values [41] provide a means of determining the contribution of different players in a cooperative game. Here they were used to determine the contribution of each input variable for predicting the output variable in each ML model.

The Shapley value analysis indicated that in some cases, only six or eight features are necessary. The six features with the highest dependence consist of $N$, $Z$, $T_Z$, and $A$ as well as one pair of shell-related features, either $\nu$ and $\zeta$, or $N_S$ and $Z_S$, depending on the ML approach. Regarding the top 8, the least influential features were often $N_E$ and $Z_E$, but there were also cases where the weakest dependence was from either pair of shell features.

A feature group containing only $N$ and $Z$ is also included to serve as a baseline. In total, seven groups of features implemented in the following analysis are shown in Table II.

## V. RESULTS AND DISCUSSION

In the following sections, the methodology used to determine the best models is discussed. In the interest of clarity in describing how well the ML models perform, we will try to share the comparison metrics for multiple data sets, including the training set, the independent testing set, as well as the entire AME 2012 and AME 2020 data sets.

### A. Determining the best models

The best model was determined for each of the three residuals using each of the ML approaches. Although the standard deviation is often the metric discussed concerning binding energy model accuracy, the standard deviations by construction emphasize the furthest outliers. When comparing nuclei from $N \geqslant 8$ and $Z \geqslant 8$, the largest deviations are often among the data for the lower mass nuclei. Instead of determining the best model based on the standard deviation, an alternative metric was chosen.

The twelve models were determined using the smallest value of the mean absolute error ($\overline{AE}$) for the independent testing data set of AME 2020 data. We are attempting to mitigate the overfitting by selecting the best models based on an independent test set. This is based on the fact that an overfit model will perform well on the training data but poorly on an independent test set. The $\overline{AE}$ values are preferable because they represent the mean absolute deviation between the model and the experimental values. So the use of $\overline{AE}$ will help mitigate a bias toward small $A$ nuclei.

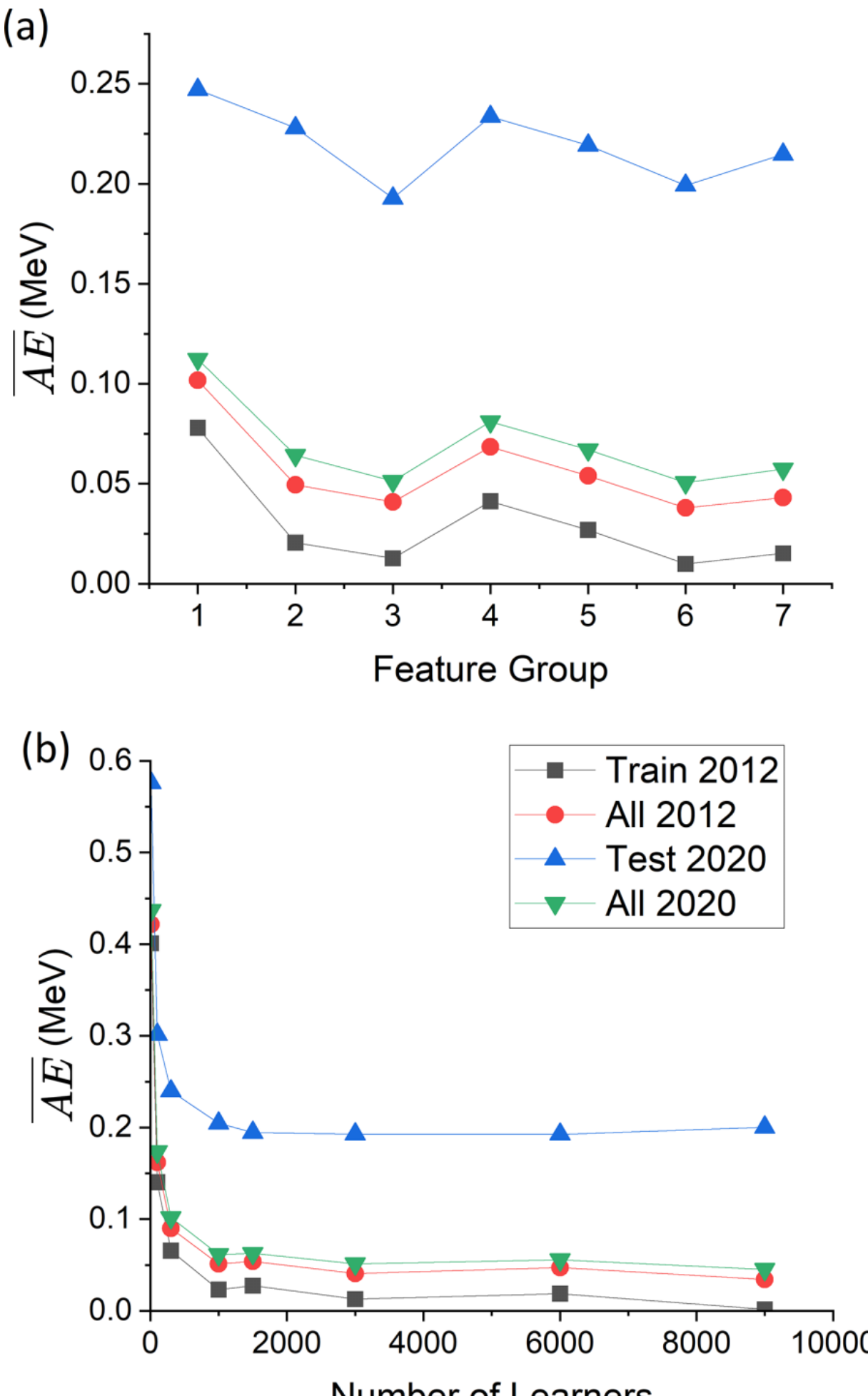

FIG. 3. The mean absolute error for $\Delta B_{\mathrm{LD6}}$ LSBET calculations (a) as a function of the feature group for 3000 learners and (b) as a number of learners for feature group 3.

Figure 3 a demonstrates why the best LD6BET model was determined to be feature group 3. In this case, the model with only six physical features outperformed models with eight or ten. This is of interest, because prior to this the literature primarily demonstrates the best model having the most features.

In general, determining the best models for SVM, GPR, and FCNN was straightforward. The $\overline{AE}$ values for all models for all feature groups were sorted and the lowest value determined the best model. For FCNN, the largest models (in both number of layers and number of nodes on those layers) were not found to provide the best results.

In the case of the LS6BET, improvement of the test set $\overline{AE}$ values did continue with the increase in model size, but the improvement was marginal. In Fig. 3(b) the number model performance for feature group 3 was plotted against the number of learners. The $\overline{AE}$ of the test set with 1000 learners was 4% above the minimum, with 1500 learners it was 1% above, and for 3000 learners it was 0.09% above the best possible model with 12 000 learners. In all three $\Delta B$ sets a cap of 3000 was imposed and the results were less than 1% above of the best values calculated.

Table III contains the model, the feature group with the lowest $\overline{AE}$ value for the test set, as as well as the comparison metrics for the training set, the entire AME 2012, and the AME 2020 set. In this table, multiple models have $\sigma$ and $\overline{AE}$ less than 100 keV, depending on the data set compared. The binding energy values for all 12 models have been included in the Supplemental Material [42].

Of the 12 best, the GPR and the LSBET-based models often outperform those from the SVM and FCNN models. The preferred model, for reasons discussed below, of the 12 best is the DZLSBET model. It was fit to the training data with both $\sigma$ and $\overline{AE}$ values below 20 keV. The DZLSBET model performed the best against the test set, with both metrics below 200 keV. The overall performance of this model compared to both the AME 2012 and AME 2020 are below 100 keV for all metrics.

### B. Understanding the models

One important feature of binding energy models is their ability to extrapolate. Figure 4 shows the resulting $\Delta B$ values for the 12 best models. The first row of this figure contains the SVM models. Our results demonstrate that the SVM will have limited reach in extrapolation. There are noticeable features near where the data was trained, but further away the value defaults to a constant value. The impact this has is that the model will predict something new for about a dozen more nuclei after the last known along an isotopic chain, but by two dozen the model will be contributing little to nothing.

TABLE III. Best trained models and corresponding evaluation metrics for $\Delta B_{\mathrm{LD5}}$, $\Delta B_{\mathrm{LD6}}$, and $\Delta B_{\mathrm{DZ}}$ using both AME 2012 [24] and AME 2020 [32] data.

| Model name | Feature group | $\sigma_{12\mathrm{Train}}$ (MeV) | $\overline{AE}_{12\mathrm{Train}}$ (MeV) | $\sigma_{12}$ (MeV) | $\overline{AE}_{12}$ (MeV) | $\sigma_{20\mathrm{Test}}$ (MeV) | $\overline{AE}_{20\mathrm{Test}}$ (MeV) | $\sigma_{20}$ (MeV) | $\overline{AE}_{20}$ (MeV) |
|---|---|---|---|---|---|---|---|---|---|
| LD5SVM | 7 | 0.097 | 0.048 | 0.150 | 0.067 | 0.352 | 0.191 | 0.182 | 0.077 |
| LD5GPR | 7 | 0.067 | 0.037 | 0.124 | 0.055 | 0.321 | 0.171 | 0.158 | 0.065 |
| LD5FCNN | 7 | 0.070 | 0.052 | 0.145 | 0.074 | 0.395 | 0.195 | 0.190 | 0.082 |
| LD5LSBET | 6 | 0.018 | 0.014 | 0.117 | 0.043 | 0.317 | 0.208 | 0.145 | 0.055 |
| LD6SVM | 5 | 0.111 | 0.049 | 0.171 | 0.071 | 0.385 | 0.196 | 0.201 | 0.080 |
| LD6GPR | 5 | 0.054 | 0.028 | 0.121 | 0.048 | 0.350 | 0.170 | 0.166 | 0.058 |
| LD6FCNN | 7 | 0.135 | 0.100 | 0.192 | 0.120 | 0.436 | 0.237 | 0.231 | 0.128 |
| LD6LSBET | 3 | 0.018 | 0.013 | 0.114 | 0.041 | 0.328 | 0.193 | 0.150 | 0.051 |
| DZSVM | 7 | 0.073 | 0.051 | 0.111 | 0.066 | 0.277 | 0.168 | 0.142 | 0.074 |
| DZGPR | 6 | 0.066 | 0.046 | 0.103 | 0.062 | 0.226 | 0.150 | 0.118 | 0.067 |
| DZFCNN | 7 | 0.119 | 0.090 | 0.151 | 0.104 | 0.303 | 0.199 | 0.173 | 0.112 |
| DZLSBET | 6 | 0.017 | 0.013 | 0.084 | 0.034 | 0.199 | 0.130 | 0.092 | 0.039 |

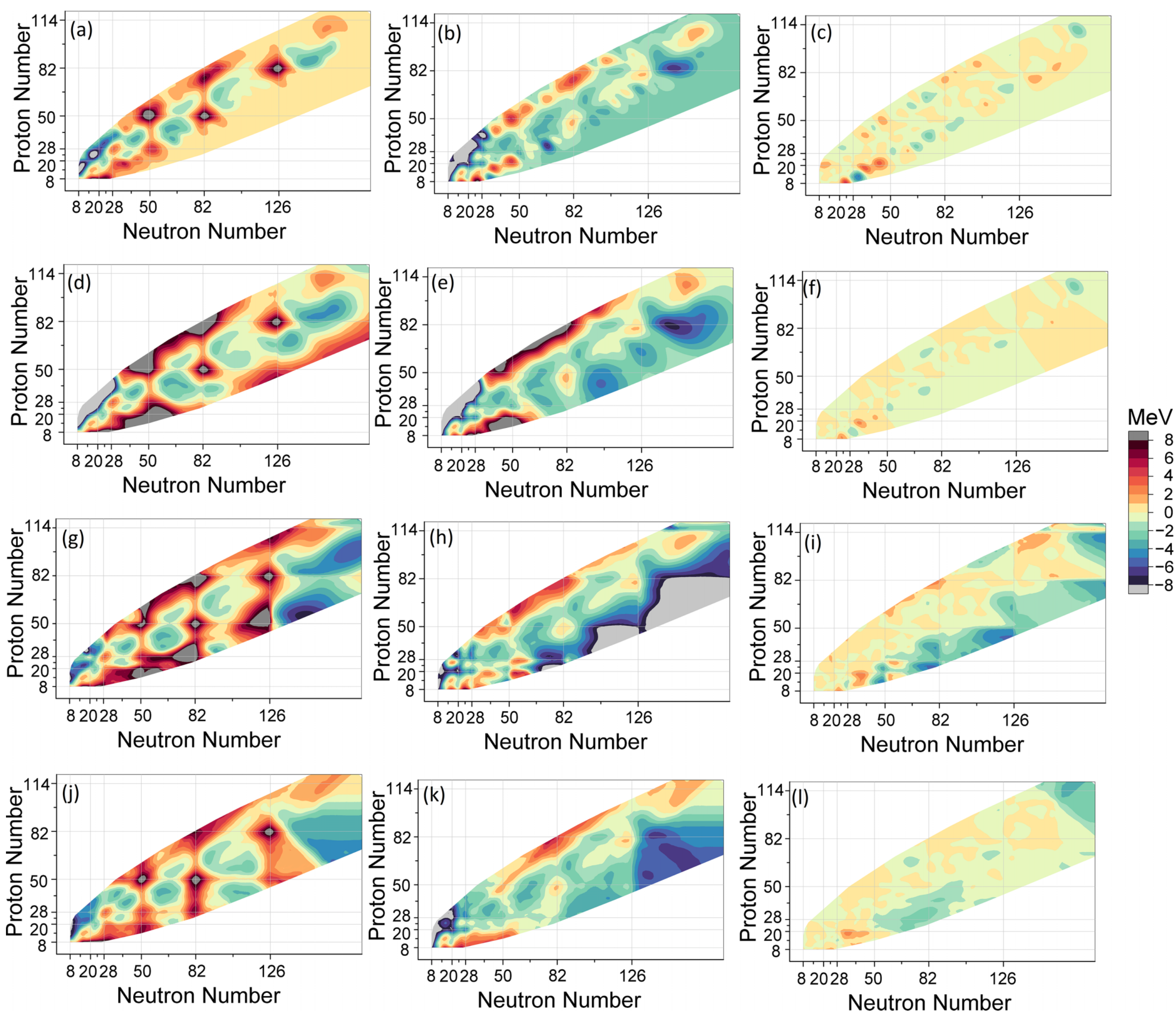

FIG. 4. The 12 best $\Delta B$ models: (a) LD5SVM with $C = 113$ and $\epsilon = 0.046$; (b) LD6SVM with $C = 382$ and $\epsilon = 0.040$; (c) DZSVM with $C = 82.1$ and $\epsilon = 0.049$; (d) LD5GPR with linear basis function and $\sigma_f = 0.00148$; (e) LD6GPR with zero basis function and $\sigma_f = 0.0496$; (f) DZGPR with zero basis function and $\sigma_f = 0.0456$; (g) LD5FCNN with a tanh activation function, two layers each consisting of 200 nodes, and $\lambda = 5.73 \times 10^{-5}$; (h) LD6FCNN with a sigmoid activation function, two layers each consisting of 100 nodes, and $\lambda = 1.39 \times 10^{-4}$; (i) DZFCNN with a tanh activation function, two layers each consisting of 100 nodes, and $\lambda = 4.98 \times 10^{-4}$; (j) LD5LSBET with 3000 learners, a minimum leaf size of 2, and $\eta = 0.102$; (k) LD6LSBET with 3000 learners, a minimum leaf size of 2, and $\eta = 0.131$; and (l) DZLSBET with 3000 learners, a minimum leaf size of 1, and $\eta = 0.0819$.

The GPR-based models similarly can have the behavior of defaulting to a constant value far from stability. In the LD5GPR and LD6GPR models, this was off the scale of the calculations, but in the DZGPR model, the occurrence was similar to the observation in the SVM models. This will be demonstrated more clearly in Sec. V B 2.

The LD5FCNN and LD6FCNN models are still good models for interpolated values, though they aren't as accurate as the others, and the extrapolation demonstrated in Fig. 4 is off scale for regions of potential astrophysical interest. The LSBET model curves, on the other hand, demonstrate that these models continue to predict new structures far from stability but on the scale with the trained values.

In the case of $\Delta B_{\mathrm{LD5}}$, the output from the models is dominated by the enhanced stability that occurs at shell closures. For $\Delta B_{\mathrm{LD6}}$ the model output is primarily a correction to the shell closure modeled by the added term in Eq. (7). For the $\Delta B_{\mathrm{DZ}}$ models, the corrections are more local and can be interpreted as corrections of a microscopic origin.

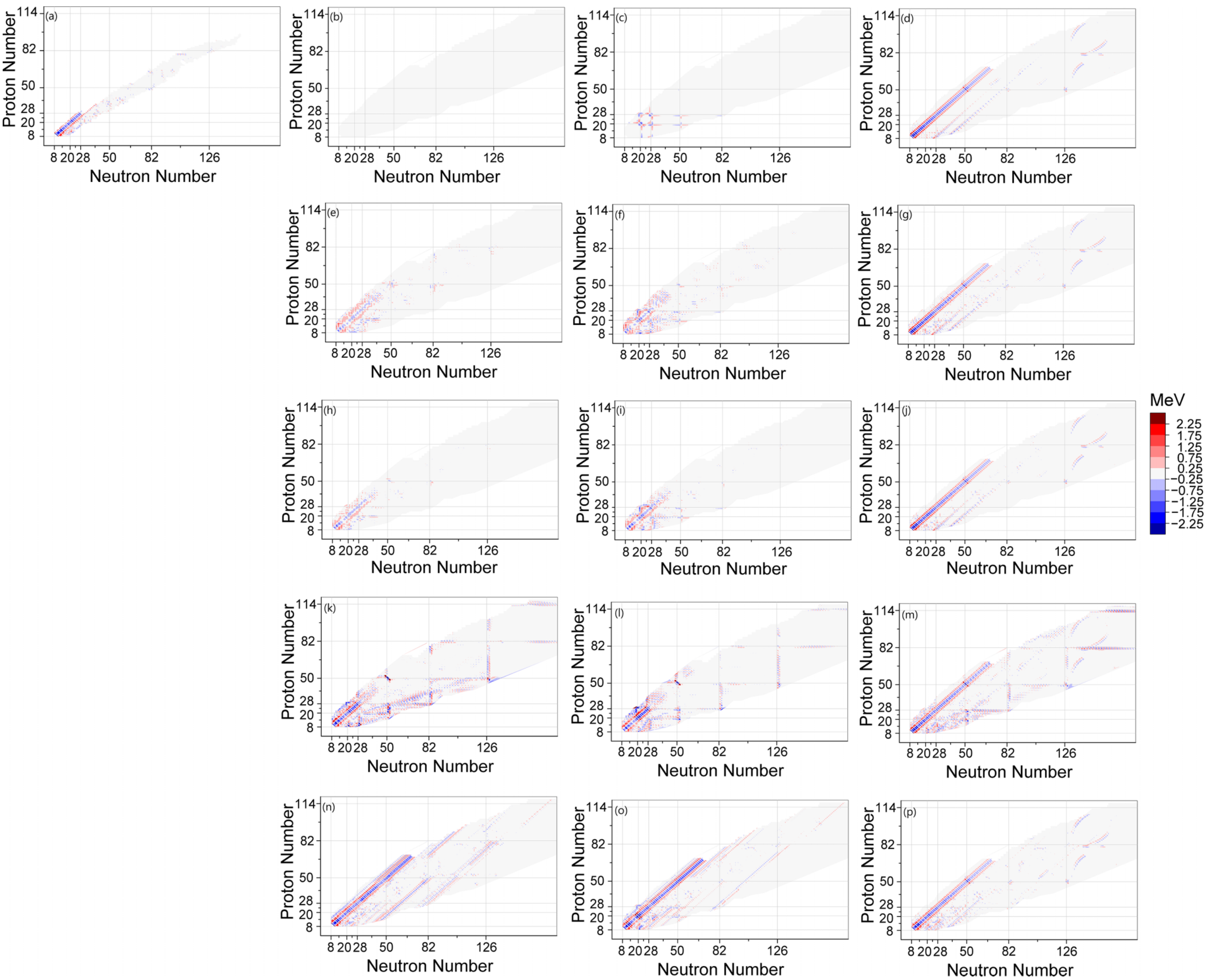

FIG. 5. The Garvey-Kelson mass relationships for experimentally measured values, the original models, and the 12 best ML models: (a) experimental values from AME 2020 [32], (b) LD5, (c) LD6, (d) DZ, (e) LD5SVM, (f) LD6SVM, (g) DZSVM, (h) LD5GPR, (i) LD6GPR, (j) DZGPR, (k) LD5FCNN, (l) LD6FCNN, (m) DZFCNN, (n) LD5LSBET, (o) LD6LSBET, and (p) DZLSBET.

#### *1. Garvey-Kelson relations*

The Garvey-Kelson mass relations [43] can provide a means of testing the relative output of a binding energy model. These relations consist of calculating a set of differences among nearby nuclei that should, in principle, add up to zero for most regions on the chart of the nuclides.

Here we will specifically use two relationships, Eqs. (1) and (3) from Ref. [44], as a feasibility check in a similar manner to Ref. [13]. Those relationships are

$$\begin{aligned} &M(Z-2,N+2)-M(Z,N)\\ &\quad+M(Z-1,N)-M(Z-2,N+1)\\ &\quad+M(Z,N+1)-M(Z-1,N+2)\approx 0, \end{aligned} \tag{13}$$

for $N \geqslant Z$, and

$$\begin{aligned} &M(Z+2,N-2)-M(Z,N)\\ &\quad+M(Z,N-1)-M(Z+1,N-2)\\ &\quad+M(Z+1,N)-M(Z+2,N-1)\approx 0, \end{aligned} \tag{14}$$

for $N < Z$.

Figure 5 demonstrates that the Garvey-Kelson relationship values are broadly found to be met by each of the best 12 models, indicating a degree that the output masses are smooth. By this metric, none of the models appear to have a substantial deviation from zero far from stability. Rather, the largest deviations are predominately among lower $A$ nuclei.

Additionally, a deviation among the Garvey-Kelson relation values noticeably occurs at $N = Z$ for all of the LSBET-based models and all of the DZ models. The same deviation is seen experimentally, as shown in Fig. 5(a). Garvey *et al.* state that the Garvey-Kelson relation holds to within 1–2 MeV for odd-odd $T_Z = 0$ nuclei and is always negative [43]. This results from enhanced pairing at $N = Z$, sometimes called the Wigner "cusp" [45]. This feature occurs in the original DZ model and it was not erased by any of the ML models, as shown in Figs. 5(b)–5(d). For the LSBET models, this attribute was generated by the ML model. The creation of these features in Garvey-Kelson relation is an indication of a benefit in using LSBET.

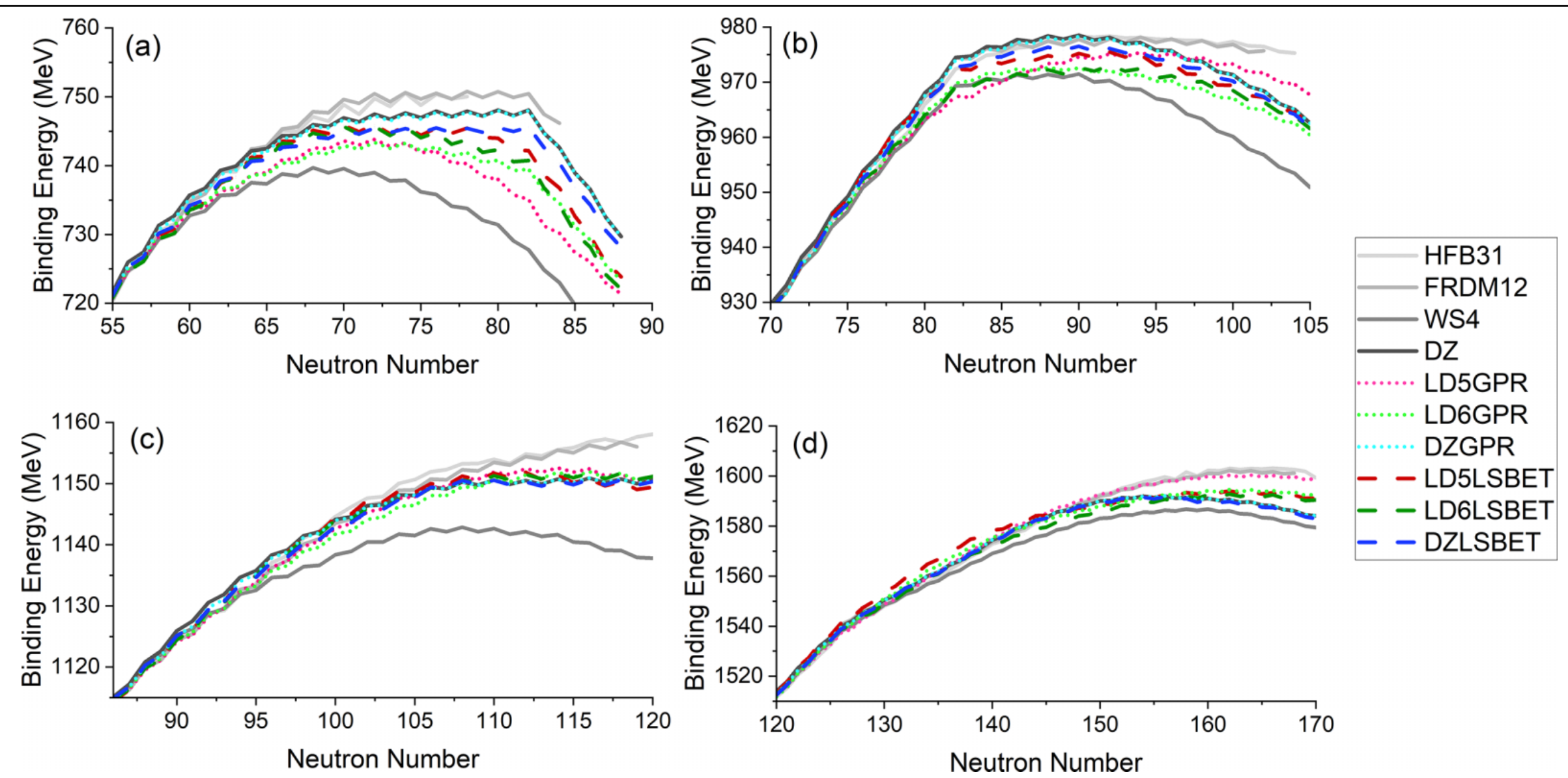

FIG. 6. Mass model extrapolation comparison for neutron-rich (a) germanium, (b) molybdenum, (c) tin, and (d) hafnium isotopes. Dotted lines denote GPR models and dashed lines indicate LSBET models. Four mass models, from Refs. [10,25,46], and [47], have been included in gray solid lines to allow for comparison.

#### 2. Comparing extrapolations

For the sake of brevity, the extrapolation capability discussion will focus on the six GPR and LSBET models. The extrapolation behavior of the models can be seen in Fig. 6. There is a spread among the GPR and LSBET models, but in every case that spread is considerably less than that seen in the four mass models included in the comparison, but the spread among these six models is still greater than the test metrics would suggest. As noted previously, the GPR- and SVM-based models all extrapolate to small values far from the training data. So essentially, those models converge to become indistinguishable from the original mass model. It is for this reason that the LSBET again is a preferred approach. It continues to make corrections to the models far from stability.

In Fig. 6(c), the tin isotopes demonstrate consistent results for the six ML models, which happen to also coincide with the DZ model. When comparing only the LSBET models (shown in dashed lines) in the neutron-rich region shown, the resulting curves can differ by up to 6 MeV for germanium, 4 MeV for molybdenum, 2 MeV for tin, and 8 MeV for hafnium. Overall, this demonstrates that these new models have less spread than say HFB31 or FRDM12 compared to WS4, and they still are not at the level of precision needed far from stability.

#### 3. Comparing with new mass excess measurements

New mass measurements continue to be made, which can serve as a second independent test set for our models. Table IV contains mass excess values from five manuscripts [48–52] describing experimental measurements that have occurred since AME 2020 [32]. All the comparisons in the listed references where the measurement was either new or differed by more than 60 keV from AME 2020 are included in Table IV.

TABLE IV. Mass excess values from new measurements and differences ($\Delta ME = ME_{\mathrm{Expt.}} - ME_{\mathrm{Model}}$), all values in keV.

| Isotope & Ref. | $ME_{\mathrm{Expt.}}$ (error) | $\Delta ME$ DZ | $\Delta ME$ HFB31 | $\Delta ME$ FRDM2012 | $\Delta ME$ WS4 | $\Delta ME$ LD5GPR | $\Delta ME$ LD5LSBET | $\Delta ME$ LD6GPR | $\Delta ME$ LD6LSBET | $\Delta ME$ DZGPR | $\Delta ME$ DZLSBET |
|---|---|---|---|---|---|---|---|---|---|---|---|
| $^{56}$V [48] | −46268(14) | −478 | 162 | −1188 | 300 | −156 | −156 | −184 | −162 | −181 | −152 |
| $^{57}$V [48] | −44371(15) | −211 | 879 | −592 | 652 | 16 | 104 | −500 | 52 | 4 | −37 |
| $^{58}$V [48] | −40361(44) | −51 | 559 | −393 | 322 | 110 | 249 | −451 | 47 | 40 | 106 |
| $^{95}$Ag [49] | −59743.3(14) | −273 | −503 | 322 | 415 | 295 | −226 | 111 | −233 | 317 | 45 |
| $^{96}$Ag [49] | −64656.69(95) | −786 | 423 | 137 | −146 | −94 | −151 | −159 | −155 | −180 | −136 |
| $^{104}$Y [50] | −53995(16) | −295 | −485 | 609 | −334 | −283 | −267 | −317 | −321 | −510 | −314 |
| $^{106}$Zr [50] | −58582.7(43) | 948 | −93 | 1311 | −162 | 22 | 123 | 133 | 158 | 391 | 81 |
| $^{109}$Nb [50] | −56810(180) | 870 | 190 | 983 | −471 | −205 | −197 | −201 | −191 | −122 | −193 |
| $^{112}$Mo [50] | −57449(118) | 1101 | 341 | 962 | 85 | −172 | 19 | −172 | 314 | −212 | 139 |
| $^{119}$Cd [51] | −84064.8(21) | 206 | −135 | −369 | 522 | −30 | −75 | −55 | −75 | −54 | −68 |
| $^{103}$Sn [52] | −67138(68) | 212 | −248 | 1023 | 83 | 58 | −686 | 95 | 102 | 148 | −73 |
| | $\sigma$ | 622 | 437 | 806 | 362 | 165 | 250 | 215 | 192 | 256 | 140 |

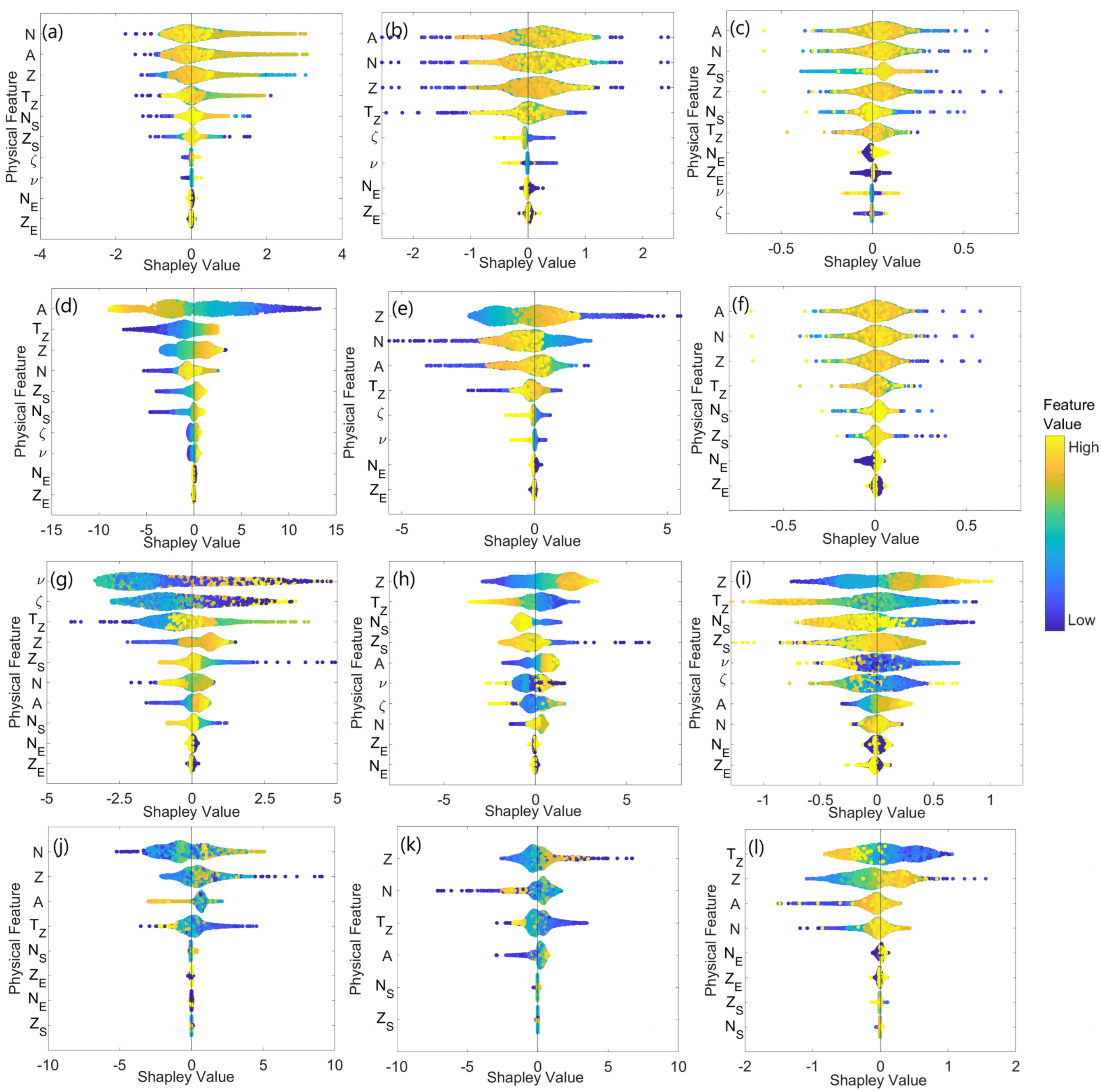

FIG. 7. Distribution of local Shapley values for the 12 best models listed in order of feature importance for (a) LD5SVM, (b) LD6SVM, (c) DZSVM, (d) LD5GPR, (e) LD6GPR, (f) DZGPR, (g) LD5FCNN, (h) LD6FCNN, (i) DZFCNN, (j) LD5LSBET, (k) LD6LSBET, and (l) DZLSBET. The vertical spread in points represents how many values are located in the same region. The predictor value color demonstrates if the value for the given input predictor was high or low. Please note that the Shapley value scale varies among the subplots.

Again, for the sake of brevity, only the mass excess values resulting from GPR and LSBET models have been included. For comparison, the difference in mass excess between the new experimental measurement following models have also been provided: Duflo Zuker (DZ) [25], Hartree Fock Bogoliubov (HFB31) [10], finite range liquid droplet (FRDM12) [46], and Weizsäcker–Skyrme (WS4) [47]. Each of these six new models outperforms the contemporary mass models. The standard deviation of the 11 new measurements compared to each model has been included at the bottom of the table.

#### 4. Shapley values

We have chosen to use Shapley values as a means of providing an explanation of the physical feature dependence for each model. Figure 7 shows the Shapley value results for all 12 models. These swarm charts show the distribution of Shapley values that are indicative of the importance of each physical feature in making predictions for each model. The vertical ordering of the physical features has been sorted in the order of importance in the model from high to low.

In all of the SVM models and the DZGPR model, the Shapley values are generally low and independent of the value of the physical feature input. This corresponds to models with complex nonlinear behavior. The LD5GPR model, on the other hand, appears to have a less complex linear behavior for most of the physical features, as indicated by clear color gradients.

The Shapley values have a broader distribution for the LSBET models in general and for the LD5GPR. This signifies that larger variance in the prediction can be achieved with these models, which, in turn, explains the better performance of these models on the test sets. In particular, DZLSBET, which outperforms all other models, shows a hierarchy of the physical variables and also a broad spread of the Shapley values for the top four physical features. Therefore, the DZLSBET can effectively model $\Delta B_{\mathrm{DZ}}$. From the lower right panel, it can be seen that $T_Z$, $Z$, $A$, and $N$ are the most important physical features in predicting $\Delta B_{\mathrm{DZ}}$, with $N_E$, $Z_E$, $Z_S$, $N_S$ being unimportant in making these predictions. This shows that, in principle, even just a four-parameter model might be sufficient.

## VI. SUMMARY AND CONCLUSION

In this work we have created a methodology for producing accurate and predictive ML models that can improve binding energy models. Four ML approaches (SVM, GPR, FCNN, and LSBET) were used to model binding energy residuals that resulted from comparing three different mass models ($\Delta B_{\mathrm{LD5}}$, $\Delta B_{\mathrm{LS6}}$, $\Delta B_{\mathrm{DZ}}$) with an AME 2012 based training set. The different ML models were trained on groups of up to 10 physical features. Here it is worth noting that nearly all of the physical features are related to the number of protons and/or neutrons in some manner.

To protect against overfitting, the metric used to evaluate model performance was the $\overline{AE}$ values determined for an independent set of 504 nuclei from the AME 2020. The $\overline{AE}$ values were preferred over the $\sigma$ values, because they treat every comparison equally, whereas the $\sigma$ values can focus on outliers and a few dozen outliers can skew the results.

In some cases, better performance was achieved with fewer physical features. In total, half of the best models contained ten physical features, while the other half used fewer than ten physical features.

SVM and GPR are techniques that are capable of providing good interpolations of data, but their ability to extrapolate is limited, and they stop contributing when taken too far from the range of the training data. FCNN and LSBET both continue to predict features that are far from stable based on the features learned elsewhere. FCNN and LSBET can be designed with variable sizes, but it is worth noting that although the accuracy of the trained models generally continues to improve with more nodes or more learners, the independent test set demonstrated convergence, and the best models required fewer nodes and fewer learners than the maximum numbers tested. Further, the best FCNN models all contained two hidden layers, outperforming larger networks with three layers.

Overall, GPR and LSBET consistently produced the best ML models. These approaches were used to generate models that resulted in $\sigma$ values of the full AME 2020 for the $\Delta B_{\mathrm{LD5}}$ and $\Delta B_{\mathrm{LD6}}$ that were the order of 150 keV. The best model produced overall that was based on $\Delta B_{\mathrm{DZ}}$ results in a $\sigma$ value of 92 keV and $\overline{AE} = 39$ keV for the full AME 2020 data set. This was confirmed by explaining the models with Shapley values.

The results from the LSBET models strike a balance of predicting new features far from stability while remaining generally on the scale with the values seen experimentally near stability. This, coupled with their strong performance regarding the fit metrics, indicates that their use should be explored further, potentially to create other models of nuclear properties.

The results discussed in this manuscript demonstrate that these new models provide improved predicted power for newly measured masses when compared to the models on which they are based and other contemporary models. These new models also potentially provide an improvement for use in astrophysics, but some other technique, perhaps involving the ensembling of multiple models, needs to be developed to produce truly reliable extrapolations.

## DATA AVAILABILITY

The data corresponding to the 12 models generated and discussed in this article are openly available [42].

[1] T. Y. Hirsh, N. Paul, M. Burkey, A. Aprahamian, F. Buchinger, S. Caldwell, J. A. Clark, A. F. Levand, L. L. Ying, S. T. Marley, G. E. Morgan, A. Nystrom, R. Orford, A. P. Galván, J. Rohrer, G. Savard, K. S. Sharma, and K. Siegl, First operation and mass separation with the CARIBU MR-TOF, Nucl. Instrum. Methods Phys. Res. Sect. B **376**, 229 (2016).

[2] M. Vilen, J. M. Kelly, A. Kankainen, M. Brodeur, A. Aprahamian, L. Canete, T. Eronen, A. Jokinen, T. Kuta, I. D. Moore, M. R. Mumpower, D. A. Nesterenko, H. Penttilä, I. Pohjalainen, W. S. Porter, S. Rinta-Antila, R. Surman, A. Voss, and J. Äystö, Precision mass measurements on neutron-rich rare-earth isotopes at JYFLTRAP: Reduced neutron pairing and implications for $r$-process calculations, Phys. Rev. Lett. **120**, 262701 (2018).

[3] M. Thoennessen, 2023 update of the discoveries of nuclides, Int. J. Mod. Phys. E **33**, 2430001 (2024).

[4] J. Wei, C. Alleman, H. Ao, B. Arend, D. Barofsky, S. Beher, G. Bollen, N. Bultman, F. Casagrande, W. Chang, Y. Choi, S. Cogan, P. Cole, C. Compton, M. Cortesi, J. Curtin, K. Davidson, S. D. Carlo, X. Du, K. Elliott *et al.*, Technological developments and accelerator improvements for the FRIB beam power ramp-up, J. Instrum. **19**, T05011 (2024).

[5] M. Mumpower, R. Surman, G. McLaughlin, and A. Aprahamian, The impact of individual nuclear properties on $r$-process nucleosynthesis, Prog. Part. Nucl. Phys. **86**, 86 (2016).

[6] S. Brett, I. Bentley, N. Paul, R. Surman, and A. Aprahamian, Sensitivity of the $r$-process to nuclear masses, Eur. Phys. J. A **48**, 184 (2012).

[7] C. F. v. Weizsäcker, Zur theorie der kernmassen, Zeitschrift für Physik **96**, 431 (1935).

[8] W. D. Myers and W. J. Swiatecki, Nuclear masses and deformations, Nucl. Phys. **81**, 1 (1966).

[9] A. Sobiczewski, Y. Litvinov, and M. Palczewski, Detailed illustration of the accuracy of currently used nuclear-mass models, At. Data Nucl. Data Tables **119**, 1 (2018).

[10] S. Goriely, N. Chamel, and J. M. Pearson, Further explorations of skyrme-Hartree-Fock-Bogoliubov mass formulas, XVI. Inclusion of self-energy effects in pairing, Phys. Rev. C **93**, 034337 (2016).

[11] A. E. Lovell, A. T. Mohan, T. M. Sprouse, and M. R. Mumpower, Nuclear masses learned from a probabilistic neural network, Phys. Rev. C **106**, 014305 (2022).

[12] L.-X. Zeng, Y.-Y. Yin, X.-X. Dong, and L.-S. Geng, Nuclear binding energies in artificial neural networks, Phys. Rev. C **109**, 034318 (2024).

[13] E. Yüksel, D. Soydaner, and H. Bahtiyar, Nuclear mass predictions using machine learning models, Phys. Rev. C **109**, 064322 (2024).

[14] X. Wu, L. Guo, and P. Zhao, Nuclear masses in extended kernel ridge regression with odd-even effects, Phys. Lett. B **819**, 136387 (2021).

[15] C.-Q. Li, C.-N. Tong, H.-J. Du, and L.-G. Pang, Deep learning approach to nuclear masses and $\alpha$-decay half-lives, Phys. Rev. C **105**, 064306 (2022).

[16] Y. Lin, J.-X. Li, and H.-F. Zhang, Transfer learning and neural networks in predicting quadrupole deformation*, Chin. Phys. C **48**, 064106 (2024).

[17] B. Lv, Z. Li, Y. Wang, and C. Petrache, Mapping low-lying states and B(E2;$0_1^+ \rightarrow 2_1^+$) in even-even nuclei with machine learning, Phys. Lett. B **857**, 139013 (2024).

[18] V. N. Vapnik, *The Nature of Statistical Learning Theory* (Springer, New York, 2000).

[19] B. Schölkopf and A. J. Smola, *Learning with Kernels: Support Vector Machines, Regularization, Optimization, and Beyond* (The MIT Press, Cambridge, MA, 2001).

[20] C. E. Rasmussen, Gaussian processes in machine learning, in *Advanced Lectures on Machine Learning: ML Summer Schools 2003, Canberra, Australia, February 2–14, 2003, Tübingen, Germany, August 4–16, 2003, Revised Lectures*, edited by O. Bousquet, U. von Luxburg, and G. Rätsch (Springer Berlin Heidelberg, 2004), pp. 63–71.

[21] C. Rasmussen and C. Williams, *Gaussian Processes for Machine Learning* (MIT Press, Cambridge, MA, 2005).

[22] M. A. Nielsen, *Neural Networks and Deep Learning* (Determination Press, San Francisco, CA, 2015), Vol. 25.

[23] L. Breiman, Random forests, Mach. Learn. **45**, 5 (2001).

[24] M. Wang, G. Audi, A. Wapstra, F. Kondev, M. MacCormick, X. Xu, and B. Pfeiffer, The AME2012 atomic mass evaluation, Chin. Phys. C **36**, 1603 (2012).

[25] J. Duflo and A. P. Zuker, Microscopic mass formulas, Phys. Rev. C **52**, R23 (1995).

[26] X.-Y. Xu, L. Deng, A.-X. Chen, H. Yang, A. Jalili, and H.-K. Wang, Improved nuclear mass formula with an additional term from the Fermi gas model, Nucl. Sci. Tech. **35**, 91 (2024).

[27] I. Bentley, Y. C. Rodríguez, S. Cunningham, and A. Aprahamian, Shell structure from nuclear observables, Phys. Rev. C **93**, 044337 (2016).

[28] J. Jänecke, T. W. O'Donnell, and V. I. Goldanskii, Isospin inversion, $n-p$ interactions, and quartet structures in $n=z$ nuclei, Phys. Rev. C **66**, 024327 (2002).

[29] I. Bentley and S. Frauendorf, Relation between Wigner energy and proton-neutron pairing, Phys. Rev. C **88**, 014322 (2013).

[30] I. Bentley, Particle-hole symmetry numbers for nuclei, Ind. J. Phys. **90**, 1069 (2016).

[31] R. Firestone, *Table of Isotopes CD-ROM* (Wiley-Interscience, Hoboken, NJ, 1999).

[32] M. Wang, W. Huang, F. Kondev, G. Audi, and S. Naimi, The AME 2020 atomic mass evaluation, (II). Tables, graphs and references*, Chin. Phys. C **45**, 030003 (2021).

[33] B. E. Boser, I. M. Guyon, and V. N. Vapnik, A training algorithm for optimal margin classifiers, in *Proceedings of the Fifth Annual Workshop on Computational Learning Theory*, COLT '92 (Association for Computing Machinery, New York, USA, 1992), pp. 144–152.

[34] H. Drucker, C. J. C. Burges, L. Kaufman, A. Smola, and V. Vapnik, Support vector regression machines, in *Advances in Neural Information Processing Systems*, edited by M. Mozer, M. Jordan, and T. Petsche (MIT Press, Cambridge, MA, 1996), Vol. 9.

[35] J. P. Janet and H. J. Kulik, *Machine Learning in Chemistry* (American Chemical Society, Washington, DC, 2020).

[36] M. Taylor, *Neural Networks: A Visual Introduction for Beginners* (Blue Windmill Media, 2017).

[37] L. Berlyand and P.-E. Jabin, *Mathematics of Deep Learning: An Introduction* (De Gruyter, Berlin, Boston, 2023).

[38] J. H. Friedman, Greedy function approximation: A gradient boosting machine, Ann. Stat. **29**, 1189 (2001).

[39] T. Hastie, R. Tibshirani, and J. Friedman, *The Elements of Statistical Learning: Data Mining, Inference, and Prediction*, Springer Series in Statistics (Springer, New York, 2009).

[40] R. F. Casten, D. S. Brenner, and P. E. Haustein, Valence $p$-$n$ interactions and the development of collectivity in heavy nuclei, Phys. Rev. Lett. **58**, 658 (1987).

[41] L. S. Shapley, *Notes on the n-Person Game, II: The Value of an n-Person Game* (Santa Monica, CA Rand Corporation, 1951).

[42] See Supplemental Material at https://www.researchgate.net/publication/388998592_12_binding_energy_models_from_PIML for the binding energy values from the 12 models.

[43] G. T. Garvey, W. J. Gerace, R. L. Jaffe, I. Talmi, and I. Kelson, Set of nuclear-mass relations and a resultant mass table, Rev. Mod. Phys. **41**, S1 (1969).

[44] J. Barea, A. Frank, J. G. Hirsch, P. V. Isacker, S. Pittel, and V. Velázquez, Garvey-Kelson relations and the new nuclear mass tables, Phys. Rev. C **77**, 041304 (2008).

[45] W. Satua, D. J. Dean, J. Gary, S. Mizutori, and W. Nazarewicz, On the origin of the Wigner energy, Phys. Lett. B **407**, 103 (1997).

[46] P. Möller, A. Sierk, T. Ichikawa, and H. Sagawa, Nuclear ground-state masses and deformations: FRDM(2012), At. Data Nucl. Data Tables **109–110**, 1 (2016).

[47] N. Wang, M. Liu, X. Wu, and J. Meng, Surface diffuseness correction in global mass formula, Phys. Lett. B **734**, 215 (2014).

[48] W. S. Porter, E. Dunling, E. Leistenschneider, J. Bergmann, G. Bollen, T. Dickel, K. A. Dietrich, A. Hamaker, Z. Hockenbery, C. Izzo, A. Jacobs, A. Javaji, B. Kootte, Y. Lan, I. Miskun, I. Mukul, T. Murböck, S. F. Paul, W. R. Plaß, D. Puentes *et al.*, Investigating nuclear structure near $n = 32$ and $n = 34$: Precision mass measurements of neutron-rich Ca, Ti, and V isotopes, Phys. Rev. C **106**, 024312 (2022).

[49] Z. Ge, M. Reponen, T. Eronen, B. Hu, M. Kortelainen, A. Kankainen, I. Moore, D. Nesterenko, C. Yuan, O. Beliuskina, L. Cañete, R. de Groote, C. Delafosse, T. Dickel, A. de Roubin, S. Geldhof, W. Gins, J. D. Holt, M. Hukkanen, A. Jaries *et al.*, High-precision mass measurements of neutron deficient silver isotopes probe the robustness of the $N = 50$ shell closure, Phys. Rev. Lett. **133**, 132503 (2024).

[50] M. Hukkanen, W. Ryssens, P. Ascher, M. Bender, T. Eronen, S. Grévy, A. Kankainen, M. Stryjczyk, O. Beliuskina, Z. Ge, S. Geldhof, M. Gerbaux, W. Gins, A. Husson, D. Nesterenko, A. Raggio, M. Reponen, S. Rinta-Antila, J. Romero, A. de Roubin *et al.*, Precision mass measurements in the zirconium region pin down the mass surface across the neutron midshell at $N$=66, Phys. Lett. B **856**, 138916 (2024).

[51] A. Jaries, M. Stryjczyk, A. Kankainen, L. Al Ayoubi, O. Beliuskina, P. Delahaye, T. Eronen, M. Flayol, Z. Ge, W. Gins, M. Hukkanen, D. Kahl, S. Kujanpää, D. Kumar, I. D. Moore, M. Mougeot, D. A. Nesterenko, S. Nikas, H. Penttilä, D. Pitman-Weymouth *et al.*, High-precision Penning-trap mass measurements of Cd and in isotopes at JYFLTRAP remove the fluctuations in the two-neutron separation energies, Phys. Rev. C **108**, 064302 (2023).

[52] Y. M. Xing, C. X. Yuan, M. Wang, Y. H. Zhang, X. H. Zhou, Y. A. Litvinov, K. Blaum, H. S. Xu, T. Bao, R. J. Chen, C. Y. Fu, B. S. Gao, W. W. Ge, J. J. He, W. J. Huang, T. Liao, J. G. Li, H. F. Li, S. Litvinov, S. Naimi *et al.*, Isochronous mass measurements of neutron-deficient nuclei from sn 112 projectile fragmentation, Phys. Rev. C **107**, 014304 (2023).